%% file: extremes.tex
\documentclass[10pt]{article}

\usepackage[english]{babel}
\usepackage[latin1]{inputenc}
\usepackage{amssymb}
\usepackage{multicol}
\usepackage[fleqn]{amsmath}
\usepackage{epsfig}
\usepackage[normalem]{ulem}
\usepackage{verbatim}
\usepackage{graphicx}
\usepackage{url} 
\usepackage{color}
\usepackage{bbm}
\usepackage{bm}
\usepackage{dsfont}
\usepackage{amsmath,amsfonts,times,latexsym,comment,times}
\usepackage{color,epsfig,rotating}

\newcommand{\bce}{\begin{center}}
\newcommand{\ece}{\end{center}}

\newcommand{\R}{I\!\!R}
\newcommand{\N}{I\!\!N}

\newcommand{\E}{\mbox{E}}

\newcommand{\1}{\mathbbm{1}}

\newcounter{cptpropo}[part]
\newenvironment{propo}[0]
{\noindent\textsc{Proposition}\,\refstepcounter{cptpropo}\thecptpropo.\it}

\newcounter{cptlemmo}[part]
\newcounter{cptexo}[part]

\title{Bayesian prior elicitation and selection for extreme values}

\author{Nicolas Bousquet\footnotemark[1]\ \footnotemark[2]
\and Merlin Keller \footnotemark[3]\ }
\footnotetext[1]{Quantmetry, 128 rue du Faubourg Saint Honoré, 75008 Paris}
\footnotetext[2]{Université Pierre-et-Marie Curie, 4 place Jussieu, 75005 Paris}
\footnotetext[3]{EDF Lab, 6 quai Watier, 78401 Chatou, France}

\begin{document}
\maketitle

\paragraph{\bf Abstract.}
A major issue of extreme value analysis is the determination of the shape parameter $\xi$ common to Generalized Extreme Value (GEV) and Generalized Pareto (GP) distributions, which drives the tail behavior, and is of major impact on the estimation of return levels and periods. Many practitioners make the choice of a Bayesian framework to conduct this assessment for accounting of parametric uncertainties, which are typically high in such analyses characterized by a low number of observations. Nonetheless, such approaches can provide large credibility domains for $\xi$, including negative and positive values, which does not allow to conclude on the nature of the tail. Considering the block maxima framework, a generic approach of the determination of the value and sign of $\xi$ arises from model selection between the Fr\'echet, Gumbel and Weibull possible domains of attraction conditionally to observations. Opposite to the common choice of the GEV as an appropriate model for {\it sampling} extreme values, this model selection must be conducted with great care. The elicitation of proper, informative and easy-to use priors is conducted based on the following principle: for all parameter dimensions they act as posteriors of noninformative priors and virtual samples. Statistics of these virtual samples can be assessed from prior predictive information, and a compatibility rule can be carried out to complete the calibration, even though they are only semi-conjugated. Besides, the model selection is conducted using a mixture encompassing framework, which allows to tackle the computation of Bayes factors. Motivating by a real case-study involving the elicitation of expert knowledge on meteorological magnitudes, the overall methodology is illustrated by toy examples too. \\

\section{Introduction}\label{sec:intro}

Since the catastrophic flood of February 1953, simultaneously in the Netherlands, England and Belgium \cite{Lamb1991}, the probabilistic assessment of many natural hazards has greatly stood over the   
extreme value statistical theory progressively established by Fr\'echet \cite{Frechet1927}, Fisher and Tippett \cite{Fisher1928} and Gumbel \cite{Gumbel1958}. The latter popularized the concepts of return periods and return levels. The use of this theory extended over the years from the study of real-world phenomena (hydrology \cite{Katz2002}; meteorology and climatology \cite{AghaKouchak2013}; maritime hydraulics \cite{Holtuijsen2010}) to many other fields, as insurance, finance \cite{Longin2016} or economics \cite{Atkinson2011ab}.  \\

\clearpage

A major feature of this theory is its twofold definition of an extreme event, in terms of {\it sampling}, which generates two statistical frameworks.  Crudely speaking, in a uni-dimensional world and under technical conditions, the maximum $X$ of an independent sample asymptotically follows an encompassing Generalized Extreme Value (GEV) distribution ({\it Block Maxima} framework, or MAXB), while the upper independent values of a sample, over a given threshold, asymptotically follows an encompassing  Generalized Pareto (GP) distribution ({\it Peaks over Threshold} framework, or POT). The various generalization of these results are well known (see \cite{Beirlant2004bb} for details) and both approaches have been proved to be equivalent \cite{Leadbetter1991}. \\

Especially, for an uni-dimensional phenomenon the two limit distributions share a common shape parameter $\xi\in\R$, the value of which severely driving the behavior of the tails of each distribution. Beyond, the assessment of return levels and periods strongly depends on the value and sign of $\xi$ \cite{Coles2001}. For $\xi>0$, the attraction domain of extreme values is Fréchet, characterized by wide return levels and no upper bound. For $\xi=0$, it is Gumbel, often encountered or postulated in hydrology \cite{Koutsoyiannis2004a}. For $\xi<0$, the attraction domain is Weibull, which has the particularity of being upperly bounded.   For this reason, numerous statisticians have proposed more and more robust estimations of extreme models \cite{Hill1975,Deheuvels1988,Dupuis1998,DellAquila2006} and refined testing procedures on $\xi$: see \cite{Chaouche2003} and \cite{Neves2008} for a review. These approaches use the GEV or GPD submodels (Fréchet, Weibull, Gumbel) rather than the encompassing GEV or GPD models themselves, in order to avoid obtaining confidence domains for$\xi$ that cover both negative and positive values.  \\

However, the large uncertainties characterizing the usual assessment of extreme values in real situations led many statistical researchers to choose a Bayesian rather than a classical framework for conducting such analysis \cite{Smith1998}. It offers the possibility to use expert knowledge about the extreme event in question, in addition to data. According to Coles \cite{Coles2001} (Chapter 9), {\it  Bayesian techniques offer an alternative that is often preferable } to classical approaches. He also notes that as the principal use of extreme value distributions is for prognosis and anticipation of future events, the most natural framework to estimate them should be \emph{predictive} \cite{Coles1996,Coles2001}. The framework needs to transfer -- as completely as possible -- uncertainty in model parameter estimation into the functions of interest (such as return periods). Bayes factors also become the natural tools of model selection.  

A complete Bayesian modeling and testing procedure between Gumbel, Fr\'echet and Weibull submodels was proposed by \cite{Parent2004} only for the POT framework, illustrated by snowfall data. In this work, informative priors were selected for their semi-conjugation properties, and calibrated based on a parametric interpretation of expert guesses (estimators of parametric quantiles). The dataset was splitted in  learning and validation subsamples, the first one being used for computing Bayes factors and the second one for testing the selected model.

However, to our knowledge, no similar work was conducted for the MAXB approach. In this framework, denoting $\theta=(\mu,\sigma,\xi)$ the usual parameterization of the submodels, the probability distribution functions (pdf) of each are:
\begin{eqnarray}
\text{Fr\'echet ${\cal{F}}(\theta)$ : } \ \ P(X<x|\theta) & = & \exp\left\{-\left(\frac{x-\mu}{\sigma}\right)^{-1/\xi}\right\} \ \ \  \label{frechet.model}\\
& & \text{with $\sigma>0$, $\xi>0$, $\mu\in\R$ and $x\geq \mu$,} \nonumber \\
\text{Weibull ${\cal{W}}(\theta)$ : } \ \ P(X<x|\theta) & = & \exp\left\{-\left(\frac{\mu-x}{\sigma}\right)^{1/\xi}\right\} \ \ \  \label{weibull.model}\\
& & \text{with $\sigma>0$, $\xi>0$, $\mu\in\R$ and $x\leq \mu$,} \nonumber \\
\text{Gumbel ${\cal{G}}_b(\theta)$ : } \ \ P(X<x|\theta) & = & \exp\left\{-\exp\left(-\frac{x-\mu}{\sigma}\right)\right\} \ \ \  \label{gumbel.model} \\
& & \text{with $\sigma>0$, $\mu\in\R$ and $x\in\R$.} \nonumber
\end{eqnarray}
This article fills this space left vacant in terms of prior modeling of $\theta$ and selection of attraction domain. \\

A particular focus is made on the elicitation of a proper prior measure $\pi(\theta)$ in non-conjugate situations (Fréchet, Weibull). Indeed, the assessment of extreme values is a public policy issue. The conservatism of public policy makers, which remain reluctant to favor Bayesian approaches to decision-helping in general \cite{Sprenger2017}, is largely due to the usual criticism of Bayesian approaches concerning the subjectivity of prior modeling, while the theoretical and practical benefits of this framework have been demonstrated for a long time \cite{Fienberg2011}, and that subjective expert knowledge always appear essential to achieve the modeling process \cite{Franklin2007}. It is besides well known that Bayes factors can strongly depend on prior effects. For these reasons, Bayesian statisticians have always to reinforce the methodologies of prior elicitation. The approach proposed here is based on exhibiting parametric prior structures that own meaningful (hyper)parameters and can be interpreted as approximate {\it posterior priors} conditional to {\it virtual data} and so-called noninformative priors. Roughly speaking, the prior hyperparameters are statistics of this virtual dataset, which can be calibrated by several means from additional information (such as expert knowledge). Such a formal approach, recommended in \cite{Wolfson2016}, allows to modulate prior information using the size of virtual data, associated to the strength of expert knowledge. It reduces to conjugate priors when the sampling model belongs to the natural exponential family (as the Gumbel model). Besides, this strategy offers the possibility, for the non-conjugate extreme models, to provide priors that offer a semi-conjugate structure.   \\ 

In addition, in the present article the approach differs from \cite{Parent2004} by several important points. First, the expert guesses are recognized as estimators of prior predictive quantiles, in accordance with the approach recommended by \cite{Kadane1998} (and beyond by many other researchers \cite{OHagan2006}) about the interpretation of anchoring values: experts do not know (even underlyingly) parameters and express knowledge on $f(x)=\E_{\pi}[f(x|\theta)]$ rather than on the probability density function $f(x|\theta)$ of each model in competition. Second, the calibration of virtual sizes is conducted between models by a rule of prior compatibility. Third, the selection of attraction domains is conducted using a new methodology of mixture modelling avoiding the computational difficulties raised by Bayes factors.

 \section{Motivating case-study} 

We consider the following case-study, taken from \cite{Ruggeri2017}. On Table \ref{corsica-values} are provided annual maxima of  pluviometry in Corsica, while expert information on this variable is summarized on Table \ref{expert-corsica}. The exchangeability and correlation between expert guesses let us interpret such information as prior predictive assessments, rather than prior parametric assessments, following the approach recommended by \cite{Kadane1998}.  \\ 

\begin{table}[hbtp]
\centering
\begin{tabular}{cccc}
     &  Pluviometry  (mm)  &  &  Pluviometry  (mm)\\
     \hline
1987       &  107.6  & 2002       &  113.2   \\
1988      &  72.4  & 2003      &  104.4   \\
1989      &  204.5  & 2004      &  66.9   \\	
1990     &  83.8   & 2005     &  136.4  \\
1991       &  142.0  & 2006      &  275.4  \\
1992       &  95.5   & 2007      &  125.0  \\
1993      &  316.1   & 2008     &  199.8  \\
1994       &  177.9  & 2009     &  51.2   \\
1995       &  87.3  & 2010      &  75.0   \\
1996      &  81.9  & 2011       &  168.2   \\
1997       &  109.1 & 2012      &  106.0   \\
1998       &  89.5  & 2013       &  72.8    \\
1999      &  150.7  & 2014      &  190.4   \\
2000      &  122.1  & 2015       &  105.0   \\
2001      &  98.2  & \\
\hline
\end{tabular}
\caption{Daily maxima per year over 2005-2011 of pluviometry  at the meteorological station of Penta-di-Casinca (Haute Corse - France). Origin: \protect\url{http://penta.meteomac.com}.}
\label{corsica-values}
\end{table}

\begin{table}[hbtp]
\centering
\begin{tabular}{cc}
Percentile order     & Pluviometry $P$ (mm) \\
 \hline
25\%    &  75   \\
50\%    &  100 \\
75\%     & 150  \\
\hline
\end{tabular}
\caption{Prior predictive information on daily maxima  per year, extrapolated by an expert from daily maxima measured at a nearby station. }
\label{expert-corsica}
\end{table}

 \section{Prior modelling} 

Prior modeling $\pi(\theta)$ for each of models (\ref{frechet.model}-\ref{gumbel.model}) is build based on the following principles:
\begin{itemize}
\item a limited number of the specifications of expert information as prior predictive percentiles should be respected;
\item the hyperparameters must have (if possible) a clear sense.
\end{itemize}
For these reasons, semi-conjugate priors are elicited in this article, which can be roughly described a  posterior priors of virtual data \cite{Wolfson2016} for a part of the parameter vector $\theta$, given usual non-informative Jeffreys priors.

\subsection{A semi-conjugate, virtual data posterior prior modeling for Fr\'echet distribution}

\subsubsection{Prior form}

Reparametrize the Fr\'echet distribution ${\cal{F}}(\theta)$ :
\begin{eqnarray*}
 P(X<x|\theta) & = & \exp\left\{-\nu\left(x-\mu\right)^{-1/\xi}\right\} 
\end{eqnarray*}
and denote now $\theta=(\mu,\nu,\xi)$ with $\nu=\sigma^{1/\xi}>0$. A nice prior form for $\pi(\theta)$ is given in next proposition. \\

\begin{propo}\label{frechet1}
Assume the Fr\'echet prior distribution $\pi(\nu,\mu,\xi)$ defined by 
\begin{eqnarray}
\nu|\mu,\xi & \sim & {\cal{G}}\left(m,s_1(\mu,\xi)\right), \nonumber \\
\xi|\mu & \sim & {\cal{IG}}\left(m,s_2(\mu)\right), \nonumber \\
\pi(\mu) & \propto & \frac{\1_{\{\mu\leq x_{e_1}\}}}{(x_{e_2}-\mu)^m s^m_2(\mu)} \label{pi.mu.frechet}
\end{eqnarray}
where $\mu<x_{e_1}< x_{e_2}$ and
\begin{eqnarray*}
s_1(\mu,\xi) & = & m(x_{e_1}-\mu)^{-1/\xi}, \\
s_2(\mu) & = & m\log\left(\frac{x_{e_2}-\mu}{x_{e_1}-\mu}\right).
\end{eqnarray*}
 Then $\pi(\nu,\mu,\xi)$  is conjugated for $\nu$ given $(\mu,\xi)$, and when $m\in\N^*$,  $\pi(\nu,\mu,\xi)=\pi^R(\nu,\mu,\xi|{\bf \tilde{x}_m})$ where $\pi^R$ is the Fr\'echet reference prior and ${\bf \tilde{x}_m}$ is a virtual Fr\'echet sample of size $m$ with statistics $\{x_{e_1},x_{e_2}\}$.   
\end{propo}

\vspace{0.25cm}


\paragraph{Remark.} We study the tails of the measure $\pi(\mu) $ defined by (\ref{pi.mu.frechet}). Denote $a=x_{e_2}-x_{e_1}>0$ and $y=x_{e_2}-\mu\in[a,\infty)$ . Then (\ref{pi.mu.frechet}) can be rewritten as proportional to
\begin{eqnarray*}
\left(y^m \log \frac{1}{1 - a/y}\right)^{-1} & \sim & \left(\log \frac{1}{1 - a/y}\right)^{-1} \ \ \ \text{when $y\to a^+$, which goes towards $0$,} \\
                                                              & \sim & \frac{1}{a y^{m-1}} \ \ \ \text{when $y\to \infty$,}
\end{eqnarray*}
by limited expansion, which goes towards $0$ if $m>1$. \\

When some finite lower bound  $\mu_{\inf}$ for $\mu$ can be assessed, $\pi(\mu)$ is proper. In this case, a simple acceptation-rejection method for sampling $\mu$ works efficiently. This approach is described in Appendix \ref{pi.mu.sampling.frechet}. The wide dispersion of $\pi(\mu)$ looks like a uniform distribution. \\

\subsubsection{Prior calibration}\label{calib.frechet}

Given a choice for $m$ and a finite lower bound $\mu_{\inf}$ for $\mu$, to calibrate hyperparameters $\omega(m)=(x_{e_1}(m),x_{e_2}(m))$ under the constraint $x_{e_1}(m)<x_{e_2}(m)$, a grid search can be used in order to minimize Cooke's criterion \cite{Cooke1991} defined by the discretization of the Kullback-Leibler loss
\begin{eqnarray}
\omega^* & = & \arg\min\limits_{\omega} \sum\limits_{i=0}^M \left(\alpha_{i+1}-\alpha_{i}\right)\log\frac{\left(\alpha_{i+1}-\alpha_{i}\right)}{\left(\tilde{\alpha}_{i+1}(\omega)-\tilde{\alpha}_{i}(\omega)\right)}, \label{cooke.regret}
\end{eqnarray}
with $\alpha_{0}=\tilde{\alpha}_0=0$ et $\alpha_{M+1}=\tilde{\alpha}_{M+1}=1$, and 
\begin{eqnarray*}
\tilde{\alpha}_{i}(\omega) & = & P(X\leq x_{\alpha}|\omega), \\ 
                            & = & \iint \left(1+\frac{\left(x_{\alpha}-\mu\right)^{-1/\xi}}{s_1(\mu,\xi)}\right)^{-m} \pi(\xi|\mu)\pi(\mu) \ d\xi d\mu.
\end{eqnarray*}
This search for $\omega(m)$ implies to estimate the double integral above. This could be done using a standard Monte Carlo approach, by sampling within the prior. However, to avoid sampling effects and get a smooth optimization, the double integral above is estimated by a constant importance sampling ($k=1,\ldots,100,000$ sampled values $(\mu_k,\xi_k)$) at each grid point:
\begin{eqnarray}
& \left.\begin{array}{lll}
\mu_k & \sim & f^{\text{\tiny IS}}_{\mu}\equiv {\cal{N}}(\kappa^{\text{\tiny IS}}_{\mu},\sigma^{\text{\tiny IS}}_{\mu}), \\
\xi_k & \sim & f^{\text{\tiny IS}}_{\xi}\equiv {\cal{IG}}\left(m,m\log\rho^{\text{\tiny IS}}_{\xi}\right)
\end{array} 
\right\} \label{is.frechet}
\end{eqnarray}

\begin{figure}[!h]
\centering
\includegraphics[scale=0.4]{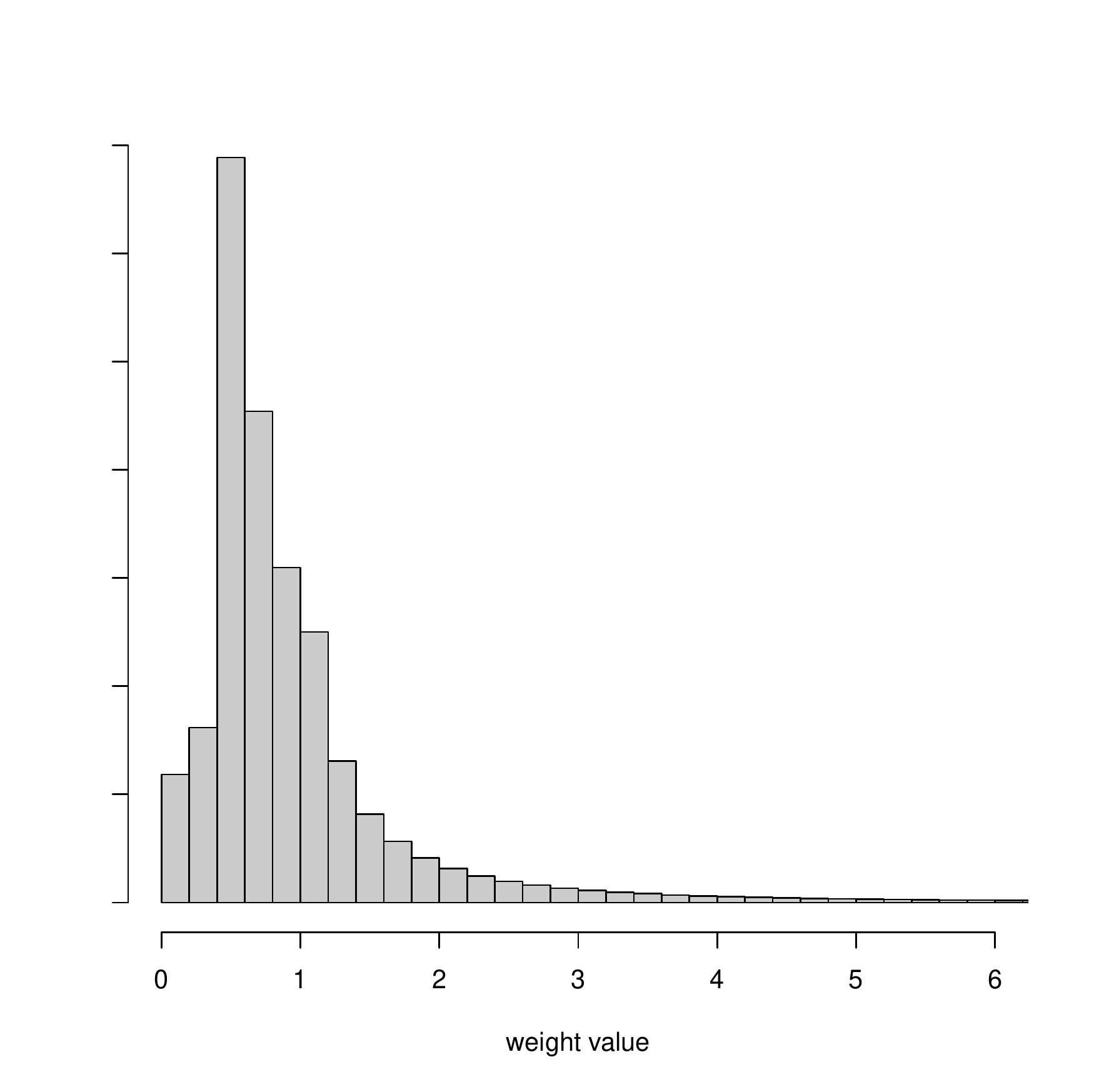}
\caption{Typical distribution of normalized importance weights $\lambda_k$ resulting from importance sampling (\ref{is.frechet}). }
\label{frechet-w}
\end{figure}

Using importance sampling in the computation of $\tilde{\alpha}_{i}(\omega)$ tackles the problem of estimating  the unknown integration constant in $\pi(\mu)$, by normalizing importance weights
\begin{eqnarray*}
\lambda_k & \propto & \frac{\pi(\mu_k)\pi(\xi_k|\mu_k)}{f^{\text{\tiny IS}}_{\mu}(\mu_k)f^{\text{\tiny IS}}_{\xi}(\xi_k)}.
\end{eqnarray*} 
using the rule $\E[\lambda_k]=1$. 
Hence the final estimation of  $\tilde{\alpha}_{i}(\omega)$ is
\begin{eqnarray*}
\hat{\alpha}_{i}(\omega) & = & M{\displaystyle \frac{\sum\limits_{k=1}^{M} \lambda_k  \left(1+\frac{\left(x_{\alpha}-\mu_k\right)^{-1/\xi_k}}{s_1(\mu_k,\xi_k)}\right)^{-m}}{\sum\limits_{k=1}^{M} \lambda_k}.}
\end{eqnarray*}
Another possibility could be to use stochastic optimization algorithms based on gradient approximation (e.g., Kiefer-Wolfowitz algorithm \cite{Kiefer1952}). Nonetheless, this basic approach led to produce values for $\omega(m)$ (Table \ref{frechet-calib}) which allows the prior predictive distribution to fit the expert guesses with a good precision. The importance sampling parameters are chosen as $\kappa^{\text{\tiny IS}}_{\mu}=0$, $\sigma^{\text{\tiny IS}}_{\mu}=50$ and  $\rho^{\text{\tiny IS}}_{\xi}=2$, after checking the balanced behavior of importance weights (see an illustration on Figure \ref{frechet-w}).\\

\begin{table}
\centering
\begin{tabular}{llll}
\hline
Virtual size $m$ & $x_{e_1}$ & $x_{e_2}$ & Order of prior predictive quartiles \\
                 &           &           &  {\footnotesize (75,100,150)}  \\
\hline
1 &  100.41 & 130.20 & $[25\%,50\%,75\%]$    \\
2 &  95.30  & 138.39 & $[24\%,49\%,74\%]$   \\
3 &  91.22  & 136.93 & $[23\%,51\%,74\%]$   \\
4 &  89.18  & 135.10 & $[24\%,50\%,74\%]$   \\
5 &  87.72  & 133.95 & $[24\%,51\%,75\%]$   \\
6 &  87.65  & 133.88 & $[24\%,50\%,75\%]$   \\
7 &  87.14  & 133.26 & $[25\%,50\%,74\%]$   \\
10 & 86.63  & 132.65 & $[25\%,51\%,75\%]$  \\
15 & 85.11  & 132.24 & $[26\%,50\%,75\%]$   \\
\hline
\end{tabular}
\caption{Hyperparameters for the Fr\'echet prior (under the constraint $\mu\geq 0$) and  computation of the effective orders  of prior predictive quartiles, which must be compared to $(25\%,50\%,75\%)$.}
\label{frechet-calib}
\end{table}

\subsection{A semi-conjugate, virtual data posterior prior modeling for Weibull distribution}

A semi-conjugate prior for Weibull can be partly established by obvious symmetry with the Fr\'echet case, adapting the ideas expressed in \cite{Bousquet2006,Bousquet2010} and \cite{Epifani2014}. Using the same parametrization, 
\begin{eqnarray*}
 P(X<x|\theta) & = & \exp\left\{-\nu\left(\mu-x\right)^{1/\xi}\right\}. 
\end{eqnarray*}

\begin{propo}\label{weibull1}
Assume the Weibull prior distribution $\pi(\nu,\mu,\xi)$ defined by 
\begin{eqnarray}
\nu|\mu,\xi & \sim & {\cal{G}}\left(m,s_3(\mu,\xi)\right), \nonumber \\
\xi|\mu & \sim & {\cal{IG}}\left(m,s_4(\mu)\right), \nonumber \\
\pi(\mu) & \propto & \frac{\1_{\{\mu\geq x_{e_4}\}}}{(\mu-x_{e_4})^m s^m_4(\mu)} \label{pi.mu.weibull}
\end{eqnarray}
where $x_{e_4}> x_{e_3}$ and
\begin{eqnarray*}
s_3(\mu,\xi) & = & m (\mu-x_{e_3})^{-1/\xi}, \\
s_4(\mu) & = & m\log\left(\frac{\mu-x_{e_3}}{\mu-x_{e_4}}\right). 
\end{eqnarray*}
 Then $\pi(\nu,\mu,\xi)$  is conjugated for $\nu$ given $(\mu,\xi)$, and when $m\in\N^*$,  $\pi(\nu,\mu,\xi)=\pi^R(\nu,\mu,\xi|{\bf \tilde{x}_m})$ where $\pi^R$ is the Weibull reference prior and ${\bf \tilde{x}_m}$ is a virtual Weibull sample of size $m$ with statistics $\{x_{e_3},x_{e_4}\}$.   
\end{propo}



\begin{figure}[!h]
\centering
\includegraphics[scale=0.6]{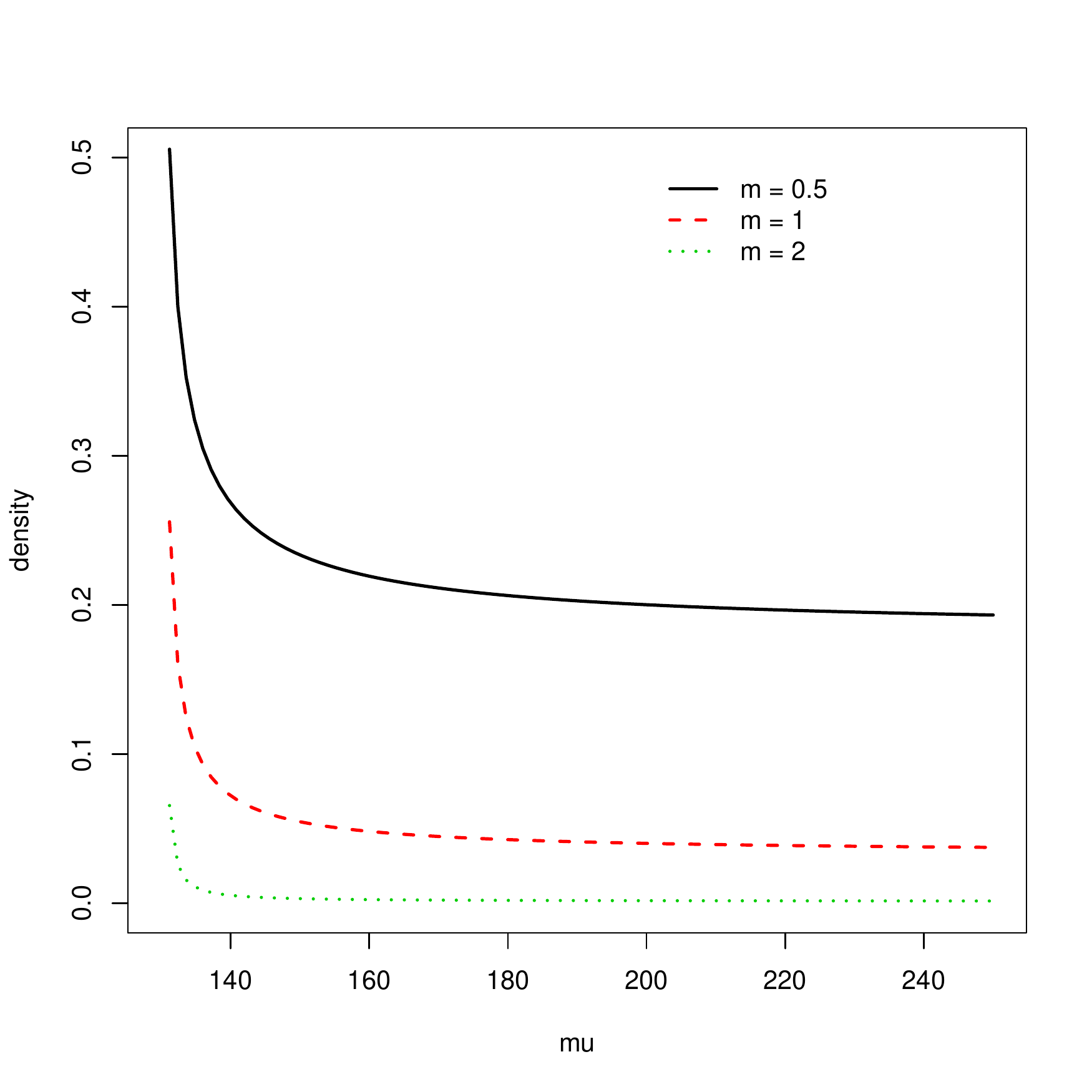}
\caption{Three prior densities $\pi(\mu)$ for the Weibull model, defined using $x_{e_3}=100$ and $x_{e_3}=130$. }
\label{prior-mu-weibull}
\end{figure}

%

\subsubsection{Prior calibration}\label{calib.weibull}

Adopting exactly the approach described at $\S$ \ref{calib.frechet}, using $\omega(m)=\{x_{e_3},x_{e_4},\rho\}$, the results of the numerical calibration are summarized on Table \ref{weibull-calib}. The results appear especially very robust about the calibration of $\rho$. \\

\begin{table}
\centering
\begin{tabular}{lllll}
\hline
Virtual size $m$ & $x_{e_3}$ & $x_{e_4}$ & $\rho$ & Order of prior predictive quartiles  \\
                 &           &           &        & {\footnotesize (75,100,150)}  \\
\hline
1 &  106.80 & 131.27 & 0.0010 & $[29\%,47\%,77\%]$  \\
2 &  96.80  & 128.00 & 0.0011 & $[28\%,49\%,76\%]$  \\
3 &  99.03  & 130.30 & 0.0010 & $[26\%,48\%,77\%]$ \\
5 &  92.74  & 128.44 & 0.0011 & $[25\%,50\%,74\%]$  \\
7 &  93.12  & 129.32 & 0.0011 & $[25\%,49\%,78\%]$  \\
10 & 86.89  & 128.45 & 0.0012 & $[27\%,50\%,77\%]$  \\
15 & 86.46  & 128.14 & 0.0011 & $[27\%,50\%,78\%]$  \\
\hline
\end{tabular}
\caption{Hyperparameters for the Weibull prior and  comparison between the effective orders of prior predictive quartiles with those of prior guesses.}
\label{weibull-calib}
\end{table}

\subsection{Conjugate prior modeling for Gumbel distribution}

Given a prior Gumbel sample $\tilde{x}_1,\ldots,\tilde{x}_{m}$ with mean $\bar{\tilde{x}}$, a conjugate prior for Gumbel distribution was elicited by \cite{Chechile2001}:
\begin{eqnarray}\label{eq:Gumbel_prior}
\pi(\mu,\sigma) & \propto & \sigma^{-m}\exp\left(m\frac{(\mu-\bar{\bf\tilde{x}}_m)}{\sigma}-\sum\limits_{i=1}^{m} \exp\left\{-\frac{\tilde{x}_i-\mu}{\sigma}\right\}\right)
\end{eqnarray} 
which is proper provided $m\geq 3$ (Theorem 1 in \cite{Chechile2001}), and 
where hyperparameters $(m,\bar{\bf\tilde{x}}_m,\tilde{x}_1,\ldots,\tilde{x}_m)$ correspond respectively to the size of a prior virtual sample, its mean and the virtual data themselves. Given true observations ${\bf x_n}$, the posterior distribution can be written as 
\begin{eqnarray*}
\pi(\mu,\sigma|{\bf x_n}) & \propto & \sigma^{-m-n}\exp\left(\{m+n\}\frac{\left(\mu-\frac{m\bar{\bf\tilde{x}}_m + n\bar{\bf x_n}}{m+n}\right)}{\sigma} \right. \\
& & \ \ \ \left.- \ \sum\limits_{i=1}^{m} \exp\left\{-\frac{\tilde{x}_i-\mu}{\sigma}\right\}-\sum\limits_{k=1}^{n} \exp\left\{-\frac{x_k-\mu}{\sigma}\right\}\right).
\end{eqnarray*}
Apart when treating historical data, prior data $\tilde{x}_i$ remains missing in practice. Nonetheless, \cite{Chechile2001} proposed to replace the missing $\tilde{x}_i$ by the quartiles of order (25\%,50\%,75\%). Doing this, by choosing $m=3$ and $\tilde{x}_1=75$, $\tilde{x}_2=100$ and $\tilde{x}_3=150$ (and consecutively $\bar{\tilde{x}}=108.33$), the effective quantile order of expert values $(75,100,150)$ take the values $(26\%,40\%,63\%)$, which remain far from the quartile orders. 

For this reason, keeping $m=3$, a grid search for $(\tilde{x}_1,\tilde{x}_2,\tilde{x}_{3})$ was conducted, under the technical constraint $\tilde{x}_1<\tilde{x}_2<\tilde{x}_3$ to avoid inopportune repetitions. The results are the following: \\
\begin{eqnarray*}
\tilde{x}_1 & = & 81, \\
\tilde{x}_2 & = & 93, \\
\tilde{x}_2 & = & 101, \\
\end{eqnarray*}
and the prior credibility orders of $(25\%,50\%,75\%)$ are correctly reproduced (up to a maximal gap of 1.5\%). \\

\begin{figure}[!h]
\centering
\includegraphics[scale=0.4]{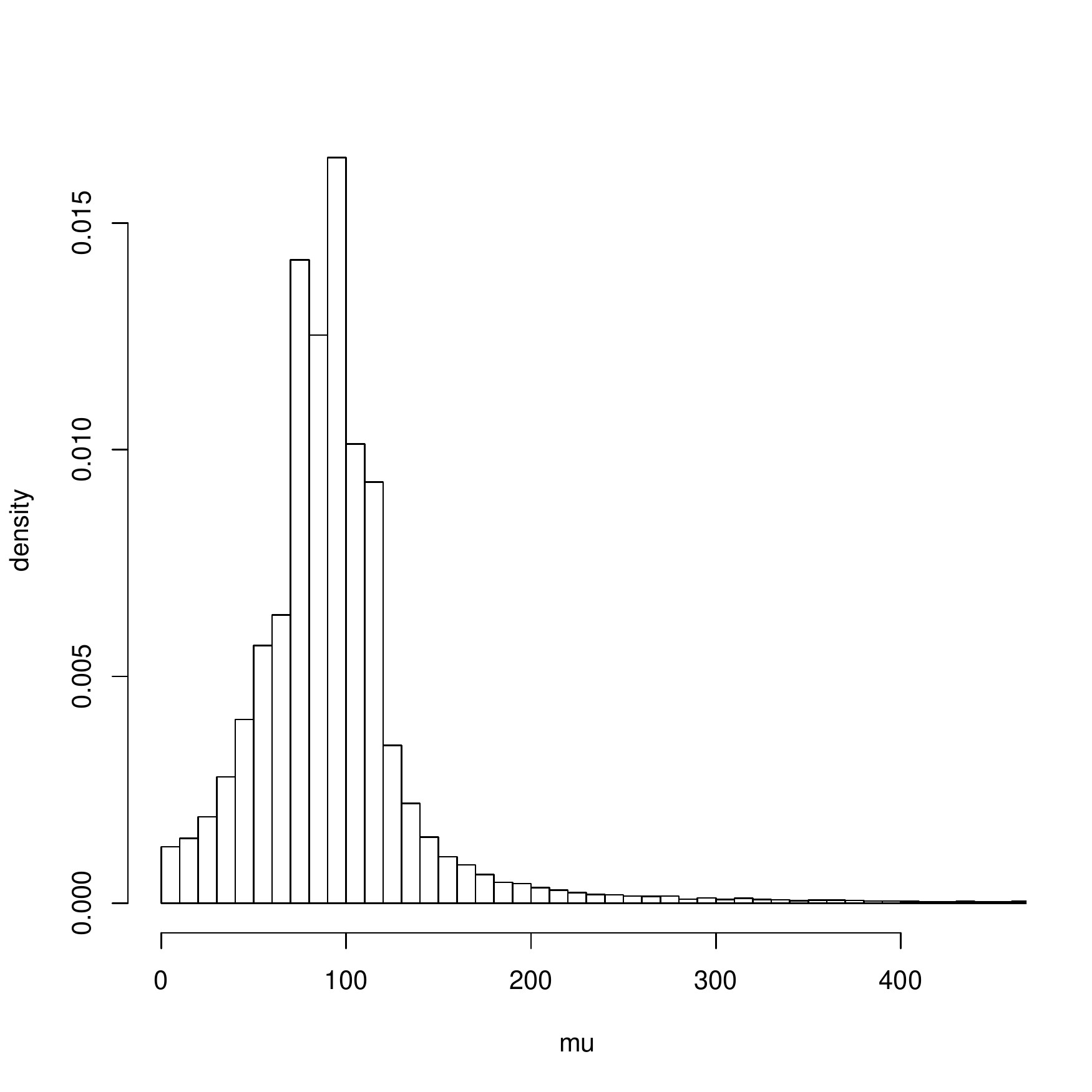}
\caption{Histogram of the marginal prior density on $\mu$.}
\label{prior-mu-gumbel}
\end{figure}

\begin{figure}[!h]
\centering
\includegraphics[scale=0.4]{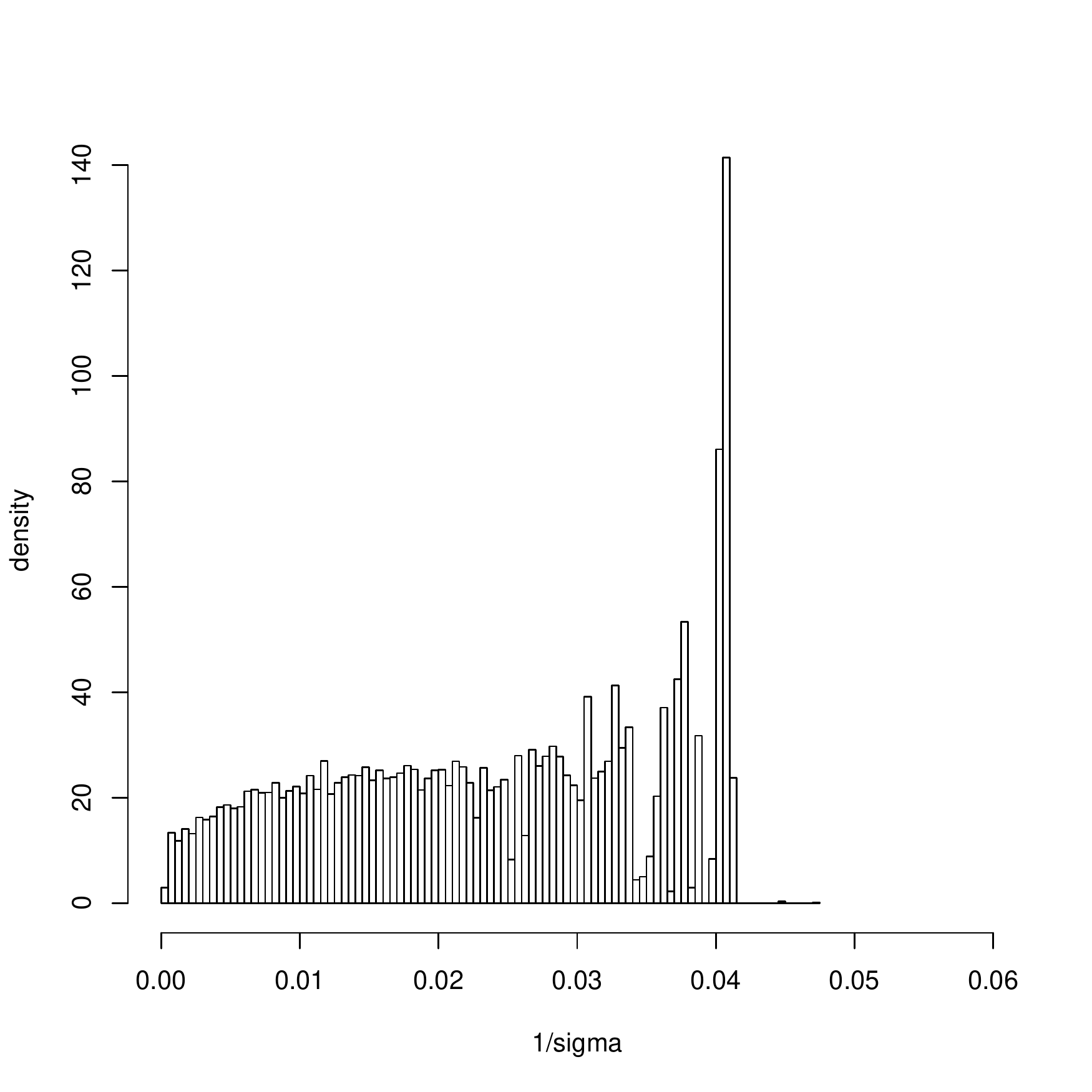}
\caption{Histogram of the marginal prior density on $1/\sigma$.}
\label{prior-inverse-sigma-gumbel}
\end{figure}

Prior simulation of $(\mu,\sigma)$ was obtained by an usual sampling importance resampling (SIR) technique based on the following importance distribution:
\begin{eqnarray*}
\mu & \sim & {\cal{E}}(1/\alpha) \ \ \ \text{with $\alpha=100$,} \\
\sigma & \sim & {\cal{IG}}\left((m-1),m\bar{\tilde{x}}\right)
\end{eqnarray*}
Given a sample $(\mu_i,\sigma_i)_i$ from this joint distribution, a prior sample can be deduced by resampling according to the (unnormalized) weights
\begin{eqnarray*}
\lambda_i & \propto & \exp\left\{\mu\left(\frac{m}{\sigma}+1/\alpha\right) - \sum\limits_{i=1}^{m} \exp\left(-\frac{\tilde{x}_i-\mu}{\sigma}\right)\right\}.
\end{eqnarray*}
See Figures \ref{prior-mu-gumbel} to \ref{prior-joint-gumbel} for a sight of marginal and correlation structures of $\pi(\mu,\sigma^{-1})$ (which is more readable graphically than $\pi(\mu,\sigma)$). \\

\begin{figure}[!h]
\centering
\caption{Joint prior structure for $(\mu,1/\sigma)$.}
\label{prior-joint-gumbel}
\end{figure}

\subsection{Balancing the prior using the virtual sizes}

To address a fair Bayesian model comparison it is recommended to get so-called compatible priors \cite{Dawid2000,Roverato2004}. Roughly speaking, when two models are competing, encoding similar prior (predictive) information on observable variable $X$ should not abusively favor a priori one of the two models, in absence of data. Several rules of compatibility were discussed in the literature \cite{Roverato2004,Celeux2006,Consonni2011} but the approach proposed by \cite{Celeux2006} has our preference, since it is based on the proximity of prior marginal distributions, which fits with the nature of the expert information considered in the motivating case-study. Briefly speaking, it stands on the fact that a parametric model $f_a$ with prior $\pi_a$ is nested in another parametric model $f_b$, accompanied by prior $\pi_b$. It is expected that the Kullback -Leibler information-loss between the more flexible marginal model $g_b$ and the constrained marginal model $g_a$:
\begin{eqnarray*}
KL\left(g_b | g_a\right) & = & \int g_b(x)\log \frac{g_b(x)}{g_a(x)} \ dx 
\end{eqnarray*} 
is minimized, where
\begin{eqnarray*}
g_c(x) & = & \int_{\Theta} g_c(x|\theta) \pi_c(\theta) \ d\theta.
\end{eqnarray*}
Gumbel can be seen as a limit case of a nested model for both Fr\'echet and Weibull models, and the virtual size $m=3$ can be fixed for this model. When fixing the virtual size for other models, each prior calibration is conducted easily. For this reason, we propose to adapt the compatibility rule proposed by \cite{Celeux2006} by operating the minimization
\begin{eqnarray}
m^*_{\Sigma} & = & \arg\min\limits_{m>0} KL\left(g_{\Sigma} | g_{{\cal{G}}_b}\right)  \label{mini.rule}
\end{eqnarray} 
where $\Sigma\in\{{\cal{F}},{\cal{W}}\}$ stands for Fr\'echet or Weibull. Since the Gumbel model is of lower dimension than Fr\'echet and Weibull, a virtual iid sample of size $m$ yields more information for Gumbel than for the two other models. Reciprocally, a similar piece of information should be distributed among the models by granting to Gumbel a lower virtual size than for the other models. This expected result is indeed obtained by solving a discretized approximation of (\ref{mini.rule}) similar to Cooke's criterion: 
\begin{eqnarray*}
m = & 5 & \text{for both the Fr\'echet and Weibull priors.} \\
\end{eqnarray*}

\section{Model selection and tail determination}

Since only proper priors are used here, Bayesian model selection and averaging can be conducted using Bayes factors \cite{Kass95,Hoeting99}. However, a novel approach for model selection, based on a mixture modeling framework, has been proposed by \cite{Kamari2014}. This has a several number of attractive features, including the fact that model estimation, selection and averaging are performed in a single algorithm, which has huge computational advantages. \cite{Keller2017} has shown that this mixture modeling formulation generalizes the classical Bayesian model selection and averaging setting, thus providing new and powerful tools to solve what has been a major challenge of computational statistics over decades.

In our case, one of the main interests in using the mixture modeling approach is that it bypasses computing the marginal likelihood, {\em i.e.} the unknown normalization constant of $\pi(\mu)$ for Fr\'echet and Weibull priors, requiring a numerical integration of the density term. The presence of negative exponential functions within this term makes this computation numerically `sensitive. Since it is wanted to minimize Monte Carlo error terms (which could be due, for instance, to two Markov chains Monte Carlo approaches -- one per non-conjugate model) that could pollute the result, the number of posterior computations is minimized.

The principle of this approach is simple. The first step consists in defining the parameter vector $\theta$ containing parameters of all candidate models. In our case, no parameters are shared between models, in spite of similar expressions for the densities, due to the fact that the prior densities turn out to be different. So the complete parameter vector is simply the concatenation of the model-specific parameter vectors, {\em i.e.} $\theta = (\theta_{\mathcal F}, \theta_{\mathcal W}, \theta_{\mathcal G_b}),$ where $\theta_{\mathcal F} = (\mu_{\mathcal F}, \nu_{\mathcal F}, \xi_{\mathcal F})$ are the Fr\'echet parameters, with prior defined by Proposition~(\ref{frechet1}), $\theta_{\mathcal W} = (\mu_{\mathcal W}, \nu_{\mathcal W}, \xi_{\mathcal W})$ are the Weibull parameters, with prior defined by Proposition~(\ref{weibull1}), and $\theta_{\mathcal G_b} = (\mu_{\mathcal G_b}, \sigma_{\mathcal G_b})$ are the Gumbel parameters, with prior defined by Equation~(\ref{eq:Gumbel_prior}).

Next, consider the mixture of all candidate models, weighted by their prior probabilities $\pi_{\mathcal M}$, such that: $\sum_{\mathcal M \in \{\mathcal F, \mathcal W,\mathcal G_b\}} \pi_{\mathcal M} = 1,$ while the likelihood of the complete dataset ${\bf x_n}$ for each model $\mathcal M$ is denoted $p_{\mathcal M}( \bf x_n| \theta_{\mathcal M})$. Hence, the mixture model likelihood is given by:
\begin{eqnarray*}
p(\bf x_n | \theta) &=& \sum_{\mathcal M \in \{\mathcal F, \mathcal W,\mathcal G_b\}}  \pi_\mathcal M \cdot p_{\mathcal M}( \bf x_n| \theta_{\mathcal M}).
\end{eqnarray*}
Hence the mixture model posterior distribution is proportional to:
\begin{eqnarray}\label{eq:mm_posterior}
\pi(\theta|\bf x_n ) &\propto& p(\bf x_n | \theta) \pi(\theta)\\
				   &\propto& \sum_{\mathcal M \in \{\mathcal F, \mathcal W,\mathcal G_b\}}\!\!\!\!\!\!\!\!  \pi_\mathcal M \cdot p_{\mathcal M}( \bf x_n| \theta_{\mathcal M}) \times\pi(\theta) \nonumber
\end{eqnarray}

The principle of mixture modeling for model comparison advocated in \cite{Keller2017} then consists in generating a sample $(\theta^{(s)})_{1\leq s\leq S}$ from this posterior, using for instance Monte-Carlo Markov chain (MCMC) algorithms \cite{Robert04}. Note that this is, in fact, the Bayesian model averaging (BMA) posterior in \cite{Hoeting99}, since, slightly modifying notations:
\begin{eqnarray}\label{eq:bma_posterior}
\pi(\theta|\bf x_n ) &\propto& \sum_{\mathcal M \in \{\mathcal F, \mathcal W,\mathcal G_b\}}\!\!\!\!\!\!\!\!  \pi_\mathcal M \cdot p_{\mathcal M}( \bf x_n| \theta) \times\pi(\theta)\nonumber\\
&\propto& \sum_{\mathcal M \in \{\mathcal F, \mathcal W,\mathcal G_b\}}\!\!\!\!\!\!\!\!  \pi_\mathcal M \cdot m_{\mathcal M}( \bf x_n) \times\pi_{\mathcal M}(\theta | \bf x_n),
\end{eqnarray}
where $m_{\mathcal M}( {\bf x_n})=\int_\theta p_{\mathcal M}( {\bf x_n}| \theta_{\mathcal M}) \pi(\theta)d\theta $ is the marginal likelihood and $\pi_{\mathcal M}(\theta | {\bf x_n}) = p_{\mathcal M}( {\bf x_n}| \theta_{\mathcal M}) \pi(\theta) / m_{\mathcal M}( {\bf x_n})$ the posterior density in model $\mathcal M$. \\

From this `posterior-averaged' sample $(\theta^{(s)})_{1\leq s\leq S}$, one can easily obtain the posterior probability for each model $\mathcal M$, as the expectation:
\begin{eqnarray*}
\mathbb P[\mathcal M | \bf x_n] &=& \int_\theta  \mathbb P[\mathcal M | \bf x_n, \theta] \pi(\theta | \bf x_n) d\theta\nonumber\\
&\widehat =& \sum_{s=1}^S \frac{\pi_\mathcal M \cdot p_{\mathcal M}( \bf x_n| \theta_{\mathcal M}^{(s)})}{p(\bf x_n | \theta^{(s)})} 
:= \sum_{s=1}^S W_{\mathcal M}(\theta^{(s)}). \\
\end{eqnarray*}
Note that the model-specific densities $p_{\mathcal M}( \bf x_n| \theta_{\mathcal M}^{(s)})$, as well as the mixture density $p( \bf x_n| \theta^{(s)})$, are usually required for computing the acceptance rate in a Metropolis-Hastings approach, so that $W_{\mathcal M}(\theta^{(s)})$ are often easily obtained generally speaking. 

Moreover, the $W_{\mathcal M}(\theta^{(s)})$ are also the weights needed to sample from model $\mathcal M$'s posterior distribution in an importance sampling approach, using the mixture model posterior distribution as an instrumental distribution, and the $\theta^{(s)}$ as proposals. \\


\section{Discussion}

\subsection{Applications in other frameworks}

These results can certaintly be useful to propose Bayesian methodologies in reliability  (especially in lifetime data analysis \cite{Rinne2008}) and physics (e.g., fracture toughness \cite{Perot2017}), where the Weibull distribution is massively used. 

\subsection{Extension to multidimensional settings}

This subsection will be finalized soon.

\section{Acknowledgments}

The authors gratefully thank Fran\c cois Gourand (M\'et\'eo-France) for having provided the dataset and his own expertise.

\addcontentsline{toc}{chapter}{Bibliographie}
\bibliographystyle{plain}
\bibliography{bibliographie}

\appendix
\input{appendix}

\end{document}

%% file: appendix.tex
\section{Proofs}

\paragraph*{Proof of Proposition \ref{frechet1}.}
Consider the following conditional prior form:
\begin{eqnarray*}
\nu|\mu,\xi & \sim & {\cal{G}}\left(m,s_1(\mu,\xi)\right), \\
\xi|\mu & \sim & {\cal{IG}}\left(m,s_2(\mu)\right)
\end{eqnarray*}
 where ${\cal{G}}(a,b)$ stands for the gamma distribution with mean $a/b$ and shape parameter $a$, and ${\cal{IG}}(a,b)$ is the corresponding inverse gamma distribution. \\

%
Given iid data ${\bf x_n}=(x_1,\ldots,x_n)$, the Fr\'echet likelihood can be written as
\begin{eqnarray}
f({\bf x_n}|\theta) & = & \frac{\nu^n}{\xi^n} \left(\prod\limits_{i=1}^n (x_i-\mu)\right)^{-1/\xi-1} \exp\left\{-\nu\sum\limits_{i=1}^n (x_i-\mu)^{-1/\xi}\right\}.\label{frechet.likeli}
\end{eqnarray}
Consequently, the conditional posterior distributions of $\nu$ and $\xi$ can be written as follows:
\begin{eqnarray*}
\nu|\mu,\xi,{\bf x_n} & \sim & {\cal{G}}\left(m+n,s_1(\mu,\xi)+\sum\limits_{i=1}^n (x_i-\mu)^{-1/\xi}\right) 
\end{eqnarray*}
and 
\begin{eqnarray*}
\pi(\xi|\mu,{\bf x_n}) & \propto & \frac{\xi^{-n-m-1}s^m_1(\mu,\xi)}{\left(s_1(\mu,\xi)+\sum\limits_{i=1}^n (x_i-\mu)^{-1/\xi}\right)^{m+n}} \exp\left\{-\frac{1}{\xi}\left(s_2(\mu) + \sum\limits_{i=1}^n \log(x_i-\mu)\right)\right\}. 
\end{eqnarray*}
Assume now that $m$ is discrete and 
\begin{eqnarray}
s_1(\mu,\xi) & = & m(x_{e_1}-\mu)^{-1/\xi}, \label{s1} \\
s_2(\mu) & = & m\log\left(\frac{x_{e_2}-\mu}{
x_{e_1}-\mu
}\right) \label{s2}
\end{eqnarray}
with  
\begin{eqnarray}
x_{e_2} & > & x_{e_1} \ > \ \mu. \label{cond.frech}
\end{eqnarray} 
Denote furthermore the shifted geometric mean of observed data
\begin{eqnarray*}
\tilde{\bar x} & = & \mu + \prod\limits_{i=1}^n (x_i-\mu)^{1/n}
\end{eqnarray*}
Then
\begin{eqnarray}
\pi(\xi|\mu,{\bf x_n}) & \propto & \frac{\xi^{-n-m-1}\exp\left\{-\frac{1}{\xi}\left(m\log(x_{e_2}-\mu) + n\log (\tilde{\bar x}-\mu) \right)\right\}}{\left(\sum\limits_{i=1}^m (x_{e_1}-\mu)^{-1/\xi} +\sum\limits_{i=1}^n (x_i-\mu)^{-1/\xi}\right)^{m+n}}. \label{xi.frech}
\end{eqnarray}  
Then $m$ plays the role of the size of a virtual Fr\'echet sample ${\bf \tilde{x}_m}=(\tilde{x}_1,\ldots,\tilde{x}_{m})$ with shifted geometric mean 
\begin{eqnarray*}
x_{e_2} & = & \mu + \prod\limits_{i=1}^m \left(\tilde{x}_i-\mu\right)^{1/m}
\end{eqnarray*}
 and shifted inverse arithmetic mean
\begin{eqnarray*}
x_{e_1} & = & \mu + \left(\sum\limits_{i=1}^m (\tilde{x}_{i}-\mu)^{-1/\xi}\right)^{-\xi},
\end{eqnarray*}
and it can be seen that
\begin{eqnarray}
\pi(\nu,\xi|\mu) & = & \pi^R(\nu,\xi|\mu,{\bf \tilde{x}_m}) \label{ref.prior.frech.1}
\end{eqnarray}
where the Fr\'echet reference prior is $\pi^{R}(\mu,\nu,\xi)\propto (\nu\xi)^{-1}$ \cite{Abbas2015}. Besides, note that
\begin{eqnarray*}
\left(\sum\limits_{i=1}^m (x_{e_1}-\mu)^{-1/\xi} +\sum\limits_{i=1}^n (x_i-\mu)^{-1/\xi}\right)^{m+n} & \geq & \frac{(m+n)^{m+n}}{ (x_{e_1}-\mu)^{m/\xi} (\tilde{\bar x}-\mu)^{n/\xi}}.
\end{eqnarray*}
Consequently, the right term in (\ref{xi.frech}) is lower than (up to a  multiplicative constant) 
\begin{eqnarray*}
\xi^{-n-m-1}\exp\left\{-\frac{s_2(\mu)}{\xi}\right\}
\end{eqnarray*}
which is the main term of a proper inverse gamma distribution. 

Notice finally that, using Bayes' rule,
\begin{eqnarray}
\pi(\mu|\nu,\xi) & \propto & \pi(\nu|\mu,\xi)\pi(\xi|\mu)\pi(\mu).\label{prior.hierar.mu}
\end{eqnarray}
Hence, from (\ref{frechet.likeli})
\begin{eqnarray*}
\pi(\mu|\nu,\xi,{\bf x_n}) & \propto & \frac{\pi(\mu) s^m_1(\mu,\xi) s^m_2(\mu)}{(\tilde{\bar x}-\mu)^{n(1+1/\xi}}\exp\left(-\nu \left[s_1(\mu,\xi) + \sum\limits_{i=1}^n (x_i-\mu)^{-1/\xi}\right] - \frac{s_2(\mu)}{\xi}\right).
\end{eqnarray*}
With
\begin{eqnarray*}
(\tilde{\bar x}-\mu)^{-n/\xi} & = & \exp\left(-\frac{n}{\xi}\log(\tilde{\bar x}-\mu)\right), \\
\exp\left(- \frac{s_2(\mu)}{\xi}\right) & = & \exp\left(-\frac{m}{\xi}\log(x_{e_2}-\mu)\right) \exp\left(\frac{m}{\xi}\log(x_{e_1}-\mu)\right),
\end{eqnarray*}
it comes
\begin{eqnarray}
\pi(\mu|\nu,\xi,{\bf x_n}) & \propto & \frac{\pi(\mu) s^m_2(\mu)}{(\tilde{\bar x}-\mu)^{n}}\exp\left(-\nu\left[\sum\limits_{i=1}^m (x_{e_1}-\mu)^{-1/\xi} + \sum\limits_{k=1}^m (x_{k}-\mu)^{-1/\xi}\right]\right) \nonumber \\
& & \ \ \ \times \ \exp\left(-\frac{1}{\xi}\left[m\log(x_{e_2}-\mu) + n\log(\tilde{\bar x}-\mu) \right]\right). \label{posterior.hierar.mu}
\end{eqnarray}
Then, from the choice of $\pi(\mu)$ given in the proposition, \\
\begin{eqnarray*}
\pi(\mu|\nu,\xi,{\bf x_n}) & \propto & {\displaystyle \frac{\exp\left(-\nu\left[\sum\limits_{i=1}^m (x_{e_1}-\mu)^{-1/\xi} + \sum\limits_{k=1}^m (x_{k}-\mu)^{-1/\xi}\right] -\frac{1}{\xi}\left[m\log(x_{e_2}-\mu) + n\log(\tilde{\bar x}-\mu) \right]\right) }{(\tilde{\bar x}-\mu)^{n} (x_{e_2}-\mu)^{m}}},
\end{eqnarray*}
and the conditional posterior is balanced between the $m$ virtual and $n$ real data, since
\begin{eqnarray*}
(\tilde{\bar x}-\mu)^{n} (x_{e_2}-\mu)^{m} & = & \prod\limits_{i=1}^{m+n} (y_i-\mu)
\end{eqnarray*}
with $y_i=x_i$ for $1\leq i \leq n$ and $y_i=\tilde{x}_{i-n}$ for $n+1\leq i \leq m+n$.  
Since $\pi^R(\mu)\propto 1$, (\ref{ref.prior.frech.1}) becomes $\pi(\nu,\xi,\mu)  =  \pi^R(\nu,\xi,\mu|{\bf \tilde{x}_m})$.  Besides, denote $y=\log(x_{e_2}-\mu)$. Then
\begin{eqnarray*}
(x_{e_2}-\mu)^{m} s^m_2(\mu) & = & \exp(m y) \left[y-\log\left(x_{e_1}-x_{e_2} + \exp(y)\right)\right]^m, \\
                             & = & \exp(m y) \left[y-y-\log\left\{1-(x_{e_2}-x_{e_1})\exp(-y)\right\}\right]^m \ \ \ \text{with $x_{e_2}>x_{e_1}$,} \\
														 & = & \exp(m y) \left[\sum\limits_{k=1}^{\infty} \frac{(x_{e_2}-x_{e_1})^k}{k!} \exp(-k y)\right]^m, \\
														 & = & \left[\sum\limits_{k=1}^{\infty} \frac{(x_{e_2}-x_{e_1})^k}{k!} \exp(-(k-1) y)\right]^m
\end{eqnarray*}
Hence
\begin{eqnarray}
\pi(\mu) & \propto & \left(1+\sum\limits_{k=1}^{\infty}\frac{1}{(k+1)!}\left(\frac{x_{e_2}-x_{e_1}}{x_{e_2}-\mu}\right)^k\right)^{-m} \ \leq \ 1. \label{sum.inf.frechet} \\
\nonumber
\end{eqnarray}

%


\paragraph*{Proof of Proposition \ref{weibull1}.}
Consider the following prior form:
\begin{eqnarray}
\nu|\mu,\xi & \sim & {\cal{G}}\left(m,s_3(\mu,\xi)\right), \nonumber\\
\xi|\mu & \sim & {\cal{IG}}\left(m,s_4(\mu)\right), \nonumber \\
\mu & \sim & \pi(\mu) \nonumber
\end{eqnarray}
The likelihood of an iid Weibull sample ${\bf x_n}$ can be written as
\begin{eqnarray*}
f({\bf x_n}|\theta) & = & \frac{\nu^n}{\xi^n} \left(\prod\limits_{i=1}^n (\mu-x_i)\right)^{1/\xi-1} \exp\left\{-\nu\sum\limits_{i=1}^n (\mu-x_i)^{1/\xi}\right\}.
\end{eqnarray*}
Remind that Berger-Bernardo's reference prior for the Weibull distribution is \cite{} $\pi^R(\mu,\nu,\xi)\propto (\nu\xi)^{-1}$. 
Consequently, the conditional posterior distributions of $\nu$ and $\xi$ can be written as follows:
\begin{eqnarray*}
\nu|\mu,\xi,{\bf x_n} & \sim & {\cal{G}}\left(m+n,s_3(\mu,\xi)+\sum\limits_{i=1}^n (\mu-x_i)^{-1/\xi}\right) 
\end{eqnarray*}
which is similar to the conditional reference posterior $\pi^R(\nu|\mu,\xi,y_1,\ldots,y_{m+n})$ of the  
sample $(y_i)_{1\leq i \leq m+n}$ defined by $y_i=x_i$ for $1\leq i \leq n$ and $y_i=\tilde{x}_{i-n}$ for $n+1\leq i \leq m+n$, where $(\tilde{x}_k)_{1\leq k \leq m}$ is a virtual Weibull sample 
with shifted arithmetic mean 
\begin{eqnarray*}
x_{e_3} & = & \mu - \left(\sum\limits_{i=1}^m (\mu-\tilde{x}_{i})^{1/\xi}\right)^{\xi}.
\end{eqnarray*}
Besides
\begin{eqnarray*}
\pi(\xi|\mu,{\bf x_n}) & \propto & \frac{\xi^{-n-m-1}s^m_3(\mu,\xi)}{\left(s_3(\mu,\xi)+\sum\limits_{i=1}^n (\mu-x_i)^{1/\xi}\right)^{m+n}} \exp\left\{-\frac{1}{\xi}\left(s_4(\mu) - \sum\limits_{i=1}^n \log(\mu-x_i)\right)\right\}. 
\end{eqnarray*}
Assume that
\begin{eqnarray}
s_3(\mu,\xi) & = & m(\mu-x_{e_3})^{1/\xi}, \label{s3} \\
s_4(\mu) & = & m\log\left(\frac{\mu-x_{e_3}}{\mu-x_{e_4}}\right) \label{s4}
\end{eqnarray}
with $x_{e_3}  <  x_{e_4}  <  \mu$.  
Denote furthermore the shifted geometric mean of observed data
\begin{eqnarray*}
\tilde{\bar x} & = & \mu - \prod\limits_{i=1}^n (\mu-x_i)^{1/n}
\end{eqnarray*}
Then
\begin{eqnarray}
\pi(\xi|\mu,{\bf x_n}) & \propto & \frac{\xi^{-n-m-1}\exp\left\{\frac{1}{\xi}\left(m\log(\mu-x_{e_4}) + n\log (\mu-\tilde{\bar x}) \right)\right\}}{\left(\sum\limits_{i=1}^m (\mu-x_{e_3})^{1/\xi} +\sum\limits_{i=1}^n (\mu-x_i)^{1/\xi}\right)^{m+n}}. \label{xi.weib}
\end{eqnarray}
Again, $m$ plays the role of the size of a virtual Weibull sample $\tilde{x}_1,\ldots,\tilde{x}_{m}$ with shifted geometric mean $x_{e_4}$ and shifted inverse arithmetic mean $x_{e_3}$, and 
$$\pi(\xi|\mu,{\bf x_n})=\pi^R(\xi|\mu,y_1,\ldots,y_{m+n}).$$ 
Notice that
\begin{eqnarray}
\left(\sum\limits_{i=1}^m (\mu-x_{e_3})^{1/\xi} +\sum\limits_{i=1}^n (\mu-x_i)^{1/\xi}\right)^{m+n} & \geq & \frac{(m+n)^{m+n}}{ (\mu-x_{e_3})^{m/\xi} (\mu-\tilde{\bar x})^{n/\xi}}. \label{propo}
\end{eqnarray}
Consequently, the right term in (\ref{xi.weib}) is lower than (up to a multiplicative constant)
\begin{eqnarray*}
\xi^{-n-m-1}\exp\left\{-\frac{s_4(\mu)}{\xi}\right\}
\end{eqnarray*}
which is the main term of a proper inverse gamma distribution. Hence $\pi(\xi|\mu,{\bf x_n})$ is proper. 

Finally, based on the same mechanisms that in (\ref{prior.hierar.mu}) and by symmetry with (\ref{posterior.hierar.mu}), using (\ref{pi.mu.weibull}) for $\pi(\mu)$, notice that
\begin{eqnarray*}
\pi(\mu|\xi,\nu,{\bf x_n}) & \propto & \frac{\1_{\{\mu\geq \max(x_{e_4},{\bf x_n})\}}}{(\mu-\tilde{\bar x})^n (\mu-x_{e_4})^m} \exp\left(-\nu\left[s_3(\mu,\xi)+\sum\limits_{i=1}^n (\mu-x_i)^{-1/\xi}\right]\right) \\
& & \ \ \times \exp\left(\frac{1}{\xi}\left[n\log(\mu-\tilde{\bar x}) + m\log(\mu-x_{e_4})\right]\right).
\end{eqnarray*}
which is exactly the conditional reference posterior of the sample $(y_i)_{1\leq i \leq m+n}$.
Notice that
\begin{eqnarray*}
\pi(\mu) & \propto & \left[1 + \sum\limits_{k=1}^{\infty} \frac{(-1)^k}{(k+1)!} \left( \frac{x_{e_4}-x_{e_3}}{\mu-x_{e_4}}\right)^k \right]^{-m}
\end{eqnarray*}
which is finite when $\mu\to\infty$ and goes to zero when $\mu\to x^+_{e_4}$. Hence $\pi(\mu)$ is proper.

\section{Sampling from $\pi(\mu)$}\label{pi.mu.sampling}

\subsection{Fr\'echet case}\label{pi.mu.sampling.frechet}

\begin{figure}[!h]
\centering
\includegraphics[scale=0.4]{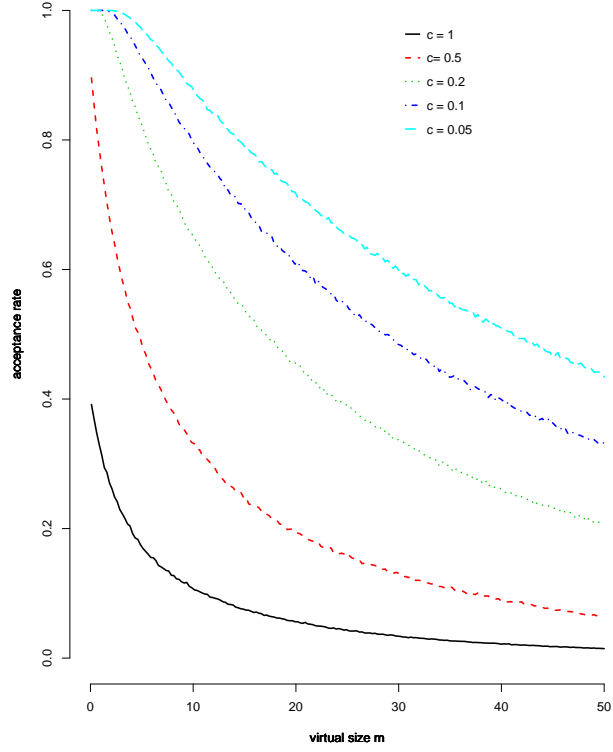}
\caption{Variations of the instrumental sampling acceptance rate using a truncated inverse gamma distribution, for several typical values of $c$. Hyperparameters $(x_{e_1},x_{e_2})$ 
are given the values $(100,130)$.}
\label{acceptance-rates}
\end{figure}

\begin{figure}[!h]
\centering
\includegraphics[scale=0.4]{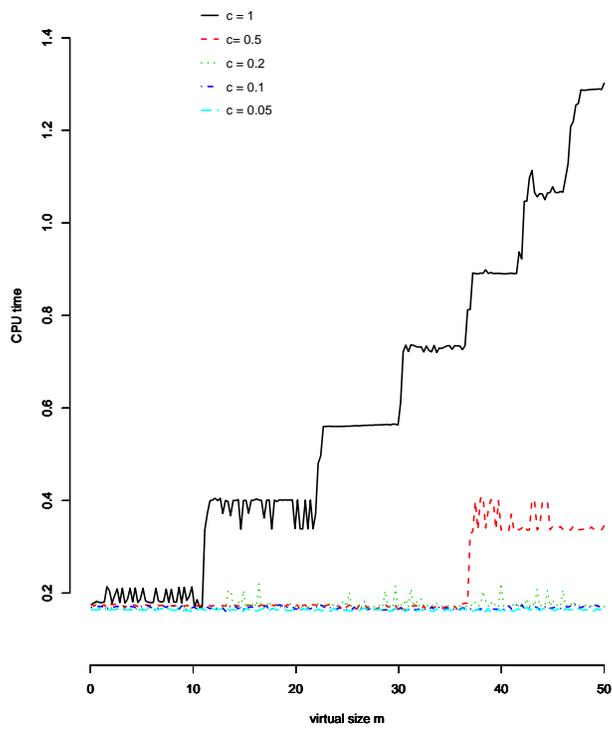}
\caption{CPU cost (in seconds) of prior sampling of $\mu$, for several typical values of $c$. Hyperparameters $(x_{e_1},x_{e_2})$ 
are given the values $(100,130)$.}
\label{cpu-cost}
\end{figure}

Assume that $\mu\geq \mu_{\inf}>-\infty$ and denote
\begin{eqnarray*}
\rho & = & \frac{x_{e_2}-x_{e_1}}{x_{e_2}-\mu_{\inf}} \in [0,1).
\end{eqnarray*}
Consider the reparametrization $$z=\left(\frac{x_{e_2}-x_{e_1}}{x_{e_2}-\mu}\right) \in [\rho,1]$$
Then $\pi(\mu)$ becomes 
\begin{eqnarray*}
\pi(z) & \propto & \tilde{\pi}(z) \ = \ \frac{z^{m-2}}{\left(-\log(1-z)\right)^m}
\end{eqnarray*}
 which is such, from (\ref{sum.inf.frechet}),
\begin{eqnarray*}
\tilde{\pi}(z) & \leq & \frac{z^{-2}}{\left(1+\frac{z}{2}\right)^m}.
\end{eqnarray*}
Hence, considering a truncated inverse gamma instrumental distribution ${\cal{IG}}(1,c)$ with support $[\rho,1]$ and density term
\begin{eqnarray*}
\pi^{\text{\tiny instr}}(z) & = & \frac{\Delta(c)}{z^{2}}\exp(-c/z)  \1_{\{\rho \leq z \leq 1\}} 
\end{eqnarray*}
with
\begin{eqnarray*}
\Delta^{-1}(c) & = & \int_{\rho}^1 \frac{\exp(-c/z)}{z^{2}} \ dz \ = \ \exp(-c) - \exp(-c/\rho).
\end{eqnarray*}
Hence
\begin{eqnarray*}
\frac{\tilde{\pi}(z)}{\pi^{\text{\tiny instr}}(z)} & \leq &  \frac{\Delta^{-1}(c)}{\left(1+\frac{z}{2}\right)^m}\exp(c/z) \ \leq \ \frac{\Delta^{-1}(c)}{\left(1+\frac{\rho}{2}\right)^m}\exp(c/\rho)
\end{eqnarray*}
As a function of $c\geq c_{\min}>0$, this upper bound is minimized in $c_{\min}$ for any couple $(m,\rho)$. For this reason, a unique value of $c=c_{\min}=0.01$ is chosen  since it allows for good acceptance rates, for a wide range of values for $m$, and a moderated CPU cost due to sampling from a truncated distribution (see an example on Figures \ref{acceptance-rates} and \ref{cpu-cost}). \\

\paragraph{Acceptation-rejection algorithm:}
\texttt{
\begin{enumerate}
\item sample $z \sim {\cal{IG}}(1,c)_{[\rho,1]}$;
\item compute $\mu=x_{e_2}-\frac{x_{e_2}-x_{e_1}}{z}$;
\item sample $u\sim {\cal{U}}[0,1]$
\item accept $\mu$ if $u\leq \Delta(c)\left(1+\frac{\rho}{2}\right)^m\exp(-c/\rho)\frac{\tilde{\pi}(z)}{\pi^{\text{\tiny instr}}(z)}$.
\end{enumerate}
}

Some representative histograms of such constrained prior distributions are plotted over Figures \ref{plot.mu.prior.examples-0} and \ref{plot.mu.prior.examples-0} for illustration, considering $\mu_{\inf}=0$ then $\mu_{\inf}=-100$. The prior distribution present a deep closeness with the uniform distribution.  \\

\begin{figure}[!h]
\centering
\includegraphics[scale=0.6]{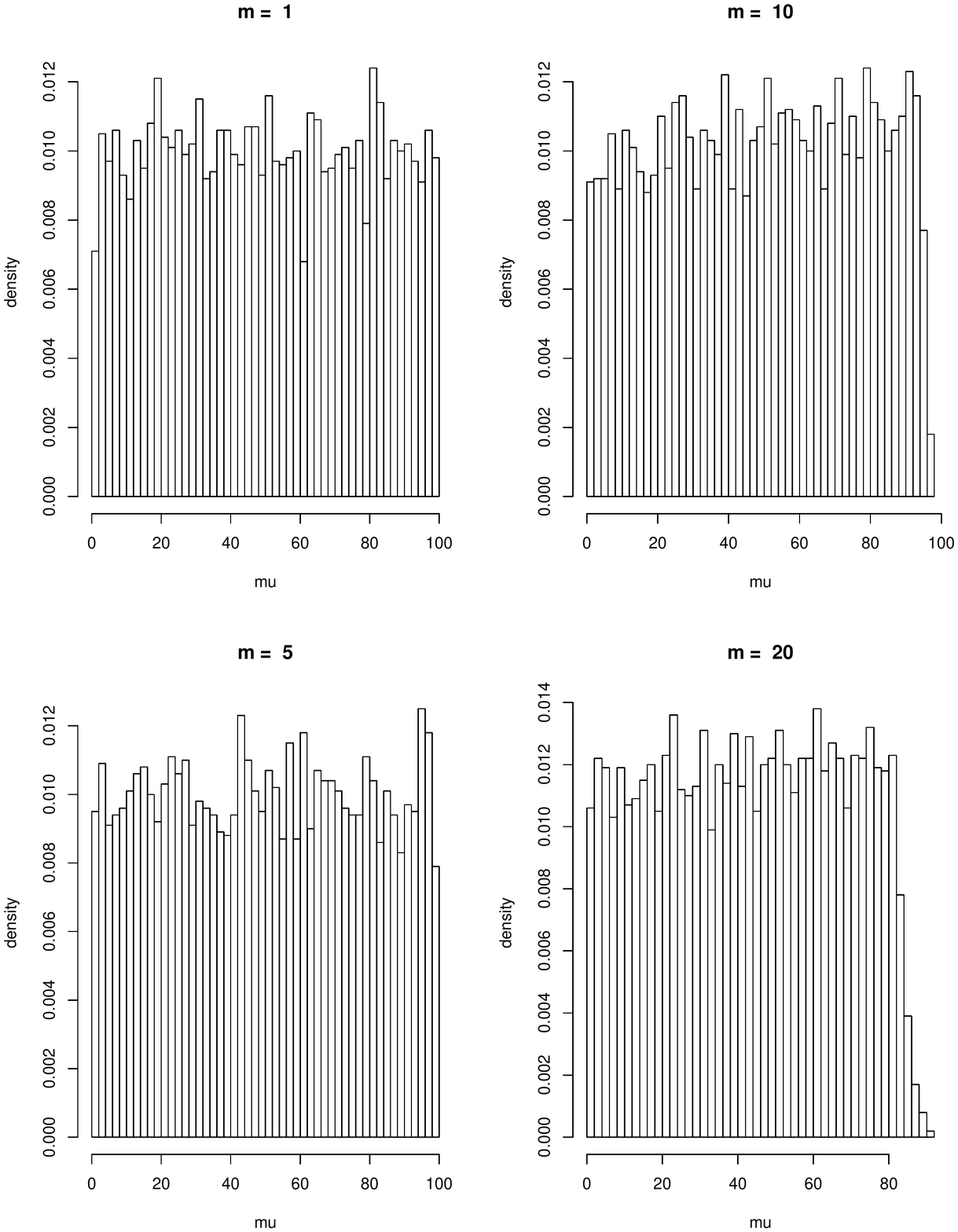}
\caption{Histograms of prior sampling for the Fr\'echet parameter  $\mu$, given various choices of $m$ and under the constraint that $\mu\geq \mu_{\inf}=0$.}
\label{plot.mu.prior.examples-0}
\end{figure}

\begin{figure}[!h]
\centering
\includegraphics[scale=0.6]{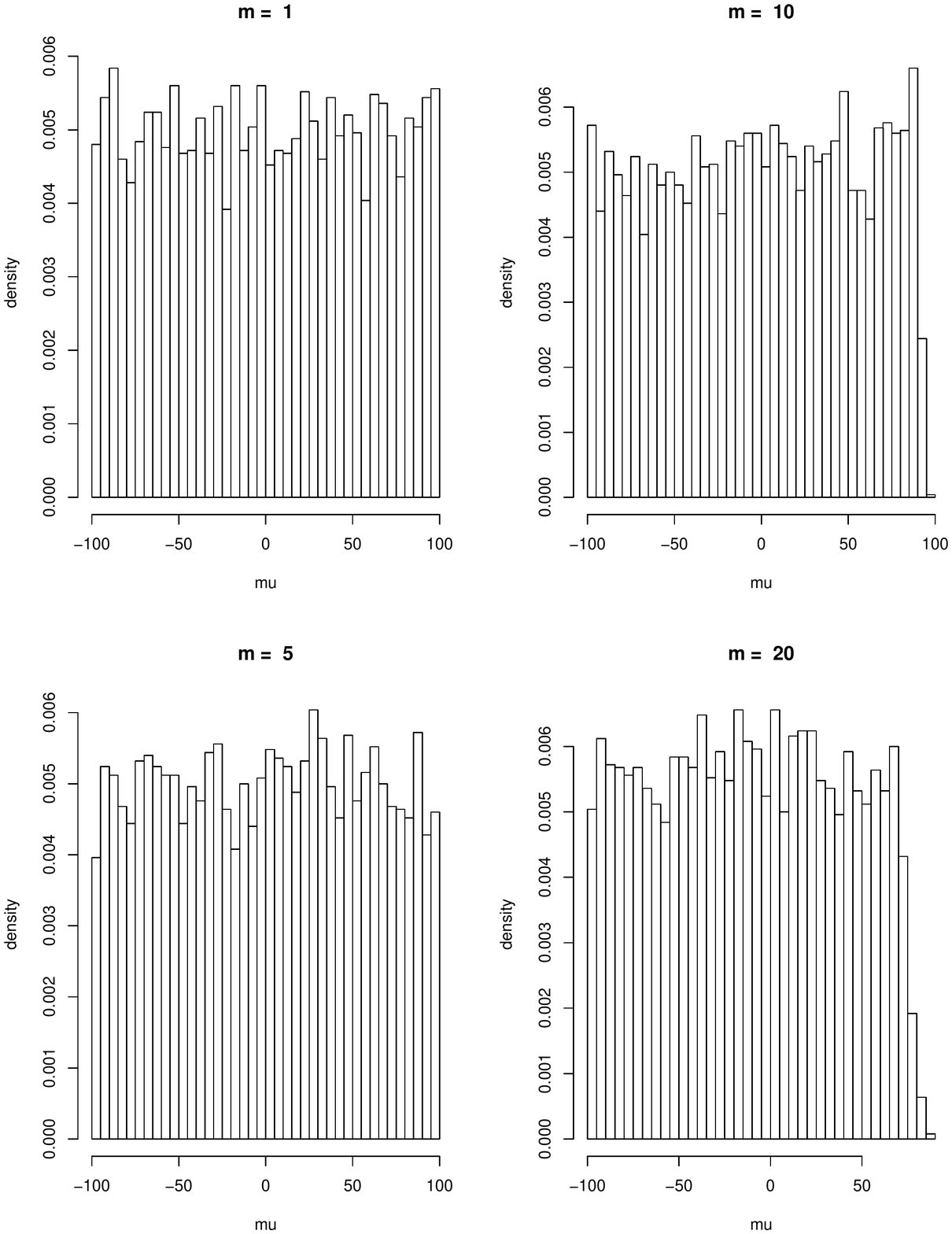}
\caption{Histograms of prior sampling for the Fr\'echet parameter  $\mu$, given various choices of $m$ and under the constraint that $\mu\geq \mu_{\inf}=-100$.}
\label{plot.mu.prior.examples-0}
\end{figure}